%% file: ifacconf.tex
\documentclass{ifacconf}

\usepackage{graphicx}      
\usepackage{natbib}        
\usepackage{amsmath}
\usepackage{pgfplots}
\usepackage{mathtools}
\usepackage{tikz}
\usepackage{enumerate}
\usepackage{microtype}
\usetikzlibrary{dsp,chains}
\definecolor{KTHmiddlegrey}{RGB}{189,188,188}
\mathtoolsset{showonlyrefs}
\usepgfplotslibrary{statistics}
\usepackage{pgfplotstable}
\usetikzlibrary{calc}
\usetikzlibrary{patterns}
\usetikzlibrary{positioning}
\usepackage{diagbox}
\usepackage[all]{kthcolors}
\pgfplotscreateplotcyclelist{kth-cycle}{%
  {thick,solid,kth-blue},
  {thick,solid,kth-red},
  {thick,solid,kth-green},
  {thick,solid,kth-yellow},
  {thick,dashed,kth-lightblue},
  {thick,dashed,kth-lightred},
  {thick,dashed,kth-lightgreen},
  {thick,solid,kth-pink},
  {thick,solid,kth-darkgray},
  {thick,solid,kth-middlegray},
  {thick,solid,kth-lightgray},
  {thick,dashed,kth-yellow},
  {thick,dashed,kth-pink},
  {thick,dashed,kth-darkgray},
  {thick,dashed,kth-middlegray},
  {thick,dashed,kth-lightgray},
}
\pgfplotscreateplotcyclelist{kth-colors}{%
  {kth-blue,fill=kth-blue},
  {kth-red,fill=kth-red},
  {kth-green,fill=kth-green},
  {kth-lightblue,fill=kth-lightblue},
  {kth-lightred,fill=kth-lightred},
  {kth-lightgreen,fill=kth-lightgreen},
  {kth-yellow,fill=kth-yellow},
  {kth-pink,fill=kth-pink},
  {kth-darkgray,fill=kth-darkgray},
  {kth-middlegray,fill=kth-middlegray},
  {kth-lightgray,fill=kth-lightgray},
}
\newlength\figureheight
\newlength\figurewidth
\newcommand{\nul}{\text{o}}
\begin{document}
\begin{frontmatter}

\title{Weighted Null-Space Fitting for Identification of Cascade Networks\thanksref{footnoteinfo}} 

\thanks[footnoteinfo]{This work was supported by the Swedish Research Council under contracts 2015-05285 and 2016-06079.k}

\author[First]{Miguel Galrinho} 
\author[Second]{Riccardo Prota}
\author[First]{Mina Ferizbegovic}
\author[First]{H\aa kan Hjalmarsson}

\address[First]{Department of Automatic Control, School of Electrical Engineering, \\KTH Royal Institute of Technology, Stockholm, Sweden \\(e-mail: \{galrinho,minafe,hjalmars\}@kth.se)}
\address[Second]{Department of Information Engineering, \\ University of Padova, Padova, Italy \\(e-mail: riccardo.prota@studenti.unipd.it)}

\begin{abstract}                
For identification of systems embedded in dynamic networks, applying the prediction error method (PEM) to a correct tailor-made parametrization of the complete network provided asymptotically efficient estimates.
However, the network complexity often hinders a successful application of PEM, which requires minimizing a non-convex cost function that in general becomes more difficult for more complex networks.
For this reason, identification in dynamic networks often focuses in obtaining consistent estimates of particular network modules of interest.
A downside of such approaches is that splitting the network in several modules for identification often costs asymptotic efficiency.
In this paper, we consider the particular case of a dynamic network with the individual systems connected in a serial cascaded manner, with measurements affected by sensor noise.
We propose an algorithm that estimates all the modules in the network simultaneously without requiring the minimization of a non-convex cost function.
This algorithm is an extension of Weighted Null-Space Fitting (WNSF), a weighted least-squares method that provides asymptotically efficient estimates for single-input single-output systems.
We illustrate the performance of the algorithm with simulation studies, which suggest that a network WNSF may also be asymptotically efficient estimates when applied to cascade networks, and discuss the possibility of extension to more general networks affected by sensor noise.
\end{abstract}

\begin{keyword}
System identification, least-squares identification, networks
\end{keyword}

\end{frontmatter}

\section{Introduction}
\input{intro.tex}

\section{Problem Statement}
\input{problem_stat.tex}

\section{Prediction Error Method}
\input{pem.tex}

\section{Weighted Null-Space Fitting}
\label{sec:wnsf}
\input{wnsf_siso.tex}

\section{Weighted Null-Space Fitting for Cascade Networks}
\label{sec:wnsfcascade}
\input{wnsf_cascade.tex}

\section{Simulations}
\label{sec:sim}
\input{simulations.tex}

\section{Discussion}
\label{sec:extensions}
\input{extensions.tex}


\bibliography{ifacconf}             
                                                   
\end{document}

%% file: intro.tex
Identification in dynamic networks gives rise to several problems and approaches.
A first problem is the detection of the network topology~\citep{materassi2010topological,materassi2012problem}.
A second problem is in determining conditions for network identifiability \citep{weerts2015identifiability,weerts2016identifiability,gevers2017identifiability}.
A third problem concerns variance analysis~\citep{Niklas:13,Niklas:14}.
Finally, there is the problem of developing appropriate methods for identification of the systems in the network, which is the problem considered in this paper.

To estimate particular modules in dynamic networks where all the nodes are observed with process noise (but no sensor noise), the inputs to every module are known exactly, and the prediction error method~\citep{ljung99} can be applied to a multi-input single-output (MISO) model~\citep{van2013identification}. 
A key issue to obtain consistent estimates is how to choose the signals that should be included in the predictor as inputs because they influence the output~\citep{dankers2016identification}.
Also, similarly to the standard closed-loop case~\citep{Forssell:99}, a noise model that is in the model set must be estimated to obtain consistent estimates; for this reason, two-stage methods~\citep{2stage} are also considered in the aforementioned works, which do not have this requirement.
If the objective is to identify the whole network, PEM can be applied with a multi-input multi-output (MIMO) model of the whole network, which with a correct parametrization provides asymptotically efficient estimates.
The main limitation with this approach is that the non-convex cost function of PEM becomes more complicated as the size of the network increases. 

If also sensor noise is present, PEM cannot be applied to estimate particular modules using internal network signals, because the inputs are noisy.
In this case, instrumental variable (IV) methods~\citep{ssiv} have been applied to provide consistent estimates of particular modules in the network~\citep{Dankers:15EIV}.
The approaches by~\citet{everitt2016identification} and~\citet{network_morsm} can also handle this case using the external excitation and auxiliary non-parametric models, and often provide more accurate estimates than IV.
In this case, because of the presence of the different noise sources, applying PEM to the whole network with a correct parametrization to obtain asymptotically efficient estimates is often a very complex problem.

For simplicity of presentation, we consider a particular case of dynamic networks: serial cascade networks, where at each note (i.e., between each module) there is either a known external excitation or a measurement affected by sensor noise (i.e., not all nodes are necessarily measured).
Although PEM can be applied to a tailor-made parametrization of this type of network, the non-convexity of the cost function is a concern.
\citet{wahlberg2009variance} have pointed out how indirect PEM~\citep{soderstrom1991indirect} can be useful to provide asymptotically efficient estimates, but only with models for which the predictor is linear in the parameters.
In some cases, subspace methods~\citep{van2012subspace} can be applied~\citep{wahlberg2008cascade,hagg2010subspace}, but they in general do not provide asymptotically efficient estimates.

The main purpose of this paper is to propose a method that provides asymptotically efficient estimates of all the cascade network modules without solving non-convex optimizations.
The proposed method is based on the Weighted Null-Space Fitting (WNSF) method, a weighted least-squares method proposed by~\citet{Galrinho:14} as an alternative to the non-convexity of PEM.
It has been shown by~\citet{wnsf_journal} that the method is consistent and asymptotically efficient for single-input single-output (SISO) systems.

In the network case, the challenge is that WNSF must incorporate the gray-box structure of the network.
In this paper, we propose an algorithm to do this.
To keep notation simple, we consider one example, and later elaborate on the extension to other networks.
In a simulation study, we illustrate the robustness of the method compared to PEM, which is prone to converge to non-global minima. 
Supported by the simulation study and the theoretical analysis in~\citet{wnsf_journal}, there are strong reasons to believe that the WNSF network extension is asymptotically efficient.

\emph{Notation.} Let $x$ be a $p$-dimensional column vector. 
Then, $\mathcal{T}_{n\times m}\{x\}$ $(n\geq p, n\geq m)$ is the $n\times m$ lower-triangular Toeplitz matrix whose first column is $[x^\top \; 0_{1\times n-p}]^\top$.

%% file: problem_stat.tex
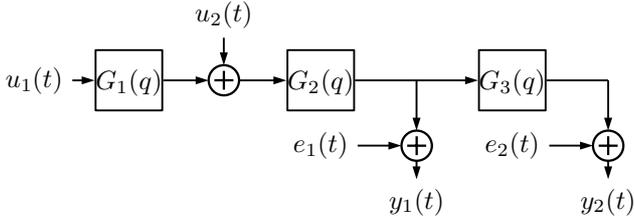
\begin{figure}
\input{network1.tex}
\vspace{-2.5em}
\caption{Serial cascade network example.}
\label{fig:network1}
\end{figure}

Consider the cascade network in Fig.~\ref{fig:network1}, where $\{G_j(q)\}_{j=1}^3$ are stable transfer functions, ($q$ is the forward-shift operator), $\{u_1(t),u_2(t)\}$ are known inputs, and $\{y_1(t),y_2(t)\}$ are measured outputs subject to mutually independent white Gaussian noises $\{e_1(t),e_2(t)\}$ with variances $\lambda_1$ and $\lambda_2$.
The relation between these signals can be written as
$y(t) = G(q) u(t) + e(t)$, where
\begin{equation}
u(t) = 
\begin{bmatrix}
u_1(t) \\ u_2(t)
\end{bmatrix}, \quad
y(t) = 
\begin{bmatrix}
y_1(t) \\ y_2(t)
\end{bmatrix}, \quad
e(t) =
\begin{bmatrix}
e_1(t) \\ e_2(t)
\end{bmatrix} ,
\end{equation}
and (argument $q$ omitted for notational simplicity)
\begin{equation}
G(q) =
\begin{bmatrix}
G_2 G_1 & G_2 \\
G_3G_2G_1 & G_3G_2
\end{bmatrix} .
\label{eq:G}
\end{equation}
For simplicity of notation, we consider this network example for the paper, and generalizations are discussed in Section~\ref{sec:extensions}.

For the model, we consider that each $G_j(q)$ is parametrized by a rational transfer function in the parameter vector $\theta_j$: 
\begin{equation}
G_j(q,\theta_j) = \frac{L_j(q,\theta_j)}{F_j(q,\theta_j)},
\label{eq:Gj}
\end{equation}
where $L_j(q,\theta_j)$ and $F_j(q,\theta_j)$ are polynomials
\begin{equation}
\begin{alignedat}{2}
F_j(q,\theta_j) = && 1 &+ f^{(j)}_1 q^{-1} + \cdots + f^{(j)}_m q^{-m}, \\
B_j(q,\theta_j) = && b^{(j)}_0 &+ b^{(j)}_1 q^{-1} + \cdots + b^{(j)}_{m-1} q^{-(m-1)},
\end{alignedat}
\label{eq:FB}
\end{equation}
and
\begin{equation}
\theta_j = 
\begin{bmatrix}
f^{(j)}_1 & \cdots & f^{(j)}_m & b^{(j)}_0 & \cdots & b^{(j)}_{m-1}
\end{bmatrix}^\top .
\label{eq:thetaj}
\end{equation}
Although each polynomials in every transfer function may have a different number of parameters, as well as number of delays in the numerator, we assume the structure in~\eqref{eq:FB} for simplicity of notation.
This gives the cascade model
\begin{equation}
y(t) = G(q,\theta) + e_t,
\label{eq:mimocascade}
\end{equation}
where
$\theta =
\begin{bmatrix}
\theta_1 & \theta_2 & \theta_3
\end{bmatrix}^\top .$
We assume also that there is $\theta_j^\nul$ such that $G_j(q)=G_j(q,\theta_j^\nul)$ and that the network is identifiable.

The problem considered in this paper is how to obtain consistent and asymptotically efficient estimates of $\theta$ without the need of minimizing a non-convex cost function.

%% file: network1.tex
\begin{tikzpicture}

	\matrix (m1) [row sep=0.8mm, column sep=3mm]
	{
		\node[coordinate]  (m00) {};     &
		\node[coordinate]          (m01) {};          &
		\node[dsp/label=above]      (m02) {$u_2(t)$};  &
		\node[coordinate]                 (m03) {};          &
        \node[coordinate]                  (m04) {};          &
		\node[coordinate]     (m05) {};  &
		\node[coordinate]                 (m06) {};\\
        \\
		\node[dsp/label=left]  (m10) {$u_1(t)$};     &
		\node[dspsquare]                  (m11) {$G_1(q)$};          &
		\node[dspadder]      (m12) {};  &
		\node[dspsquare]   (m13) {$G_2(q)$};          &
        \node[coordinate]                  (m14) {};          &
		\node[dspsquare]     (m15) {$G_3(q)$};  &
		\node[coordinate]  (m16) {};    \\
        \\
		\node[coordinate]  (m20) {};     &
		\node[coordinate]                  (m21) {};          &
		\node[coordinate]      (m22) {};  &
		\node[dsp/label=left]   (m23) {$e_1(t)$};          &
        \node[dspadder]   (m24) {};          &
		\node[dsp/label=left]     (m25) {$e_2(t)$};  &
		\node[dspadder]  (m26) {};    \\
        \\
		\node[coordinate]  (m30) {};     &
		\node[coordinate]                  (m31) {};          &
		\node[coordinate]      (m32) {};  &
		\node[coordinate]   (m33) {};          &
        \node[dsp/label=below]                  (m34) {$y_1(t)$};          &
		\node[coordinate]     (m35) {};  &
		\node[dsp/label=below]  (m36) {$y_2(t)$};     \\
        \\
	};

    
    \begin{scope}[start chain]
		\chainin (m10);
		\chainin (m11) [join=by dspconn];
	\end{scope}
    \begin{scope}[start chain]
		\chainin (m11);
		\chainin (m12) [join=by dspconn];
	\end{scope}
    \begin{scope}[start chain]
		\chainin (m12);
		\chainin (m13) [join=by dspconn];
	\end{scope}
    \begin{scope}[start chain]
		\chainin (m02);
		\chainin (m12) [join=by dspconn];
	\end{scope}
    \begin{scope}[start chain]
		\chainin (m13);
		\chainin (m15) [join=by dspconn];
	\end{scope}
    \begin{scope}[start chain]
		\chainin (m14);
		\chainin (m24) [join=by dspconn];
	\end{scope}
    \begin{scope}[start chain]
    	\chainin (m15);
		\chainin (m16) [join=by dspline];
        \chainin (m26) [join=by dspconn];
	\end{scope}
    \begin{scope}[start chain]
		\chainin (m23);
		\chainin (m24) [join=by dspconn];
	\end{scope}
        \begin{scope}[start chain]
		\chainin (m25);
		\chainin (m26) [join=by dspconn];
	\end{scope}
    \begin{scope}[start chain]
		\chainin (m24);
		\chainin (m34) [join=by dspconn];
	\end{scope}
    \begin{scope}[start chain]
		\chainin (m26);
		\chainin (m36) [join=by dspconn];
	\end{scope}
    

\end{tikzpicture}

%% file: pem.tex
In this section, we review how the prediction error method (PEM) can be applied to the cascade network in Fig.~\ref{fig:network1} to obtain asymptotically efficient estimate of $\theta$.
We discuss the limitation of non-convexity and how initial estimates can be obtained.
This limitation serves as motivation for the method we will propose. 

The idea of PEM is to minimize a cost function of the prediction errors.
For the cascade network considered in this paper, the prediction errors are given by
\begin{equation}
\varepsilon(t,\theta) = y(t) - G(q,\theta),
\end{equation}
where $\varepsilon(t,\theta)$ is white noise sequence.
Then, PEM with a quadratic cost consists of minimizing
\begin{equation}
V_N(\theta) = \det \left[ \frac{1}{N} \sum_{t=1}^N \varepsilon(t,\theta) \varepsilon^\top(t,\theta) \right],
\label{eq:Vpem}
\end{equation}
where $N$ is the sample size.

The global minimizer $\hat{\theta}_N$ of this cost function is an asymptotically efficient estimate of $\theta$.
The problem is that this is, in general, a non-convex cost function, for which optimization algorithms can converge to a non-global minimum.
Although this is the case also for SISO systems, the cascade case is of more concern.
First, the size of the problem increases with the number of systems in the network; second, methods to provide initialization points for PEM may not be directly applicable to cascade networks.

Nevertheless, the following straightforward procedure could be applied to provide initial estimates for the cost function~\eqref{eq:Vpem}.
First, estimate an over-parametrized MIMO OE model with
\begin{equation}
G(q,\eta) = 
\begin{bmatrix}
G_{11}(q,\eta_{11}) & G_{12}(q,\eta_{12}) \\
G_{21}(q,\eta_{21}) & G_{22}(q,\eta_{22}) 
\end{bmatrix},
\label{eq:mimo_oe_overp}
\end{equation}
where this parametrization is defined similarly to~\eqref{eq:Gj}, \eqref{eq:FB} and \eqref{eq:thetaj}.
The difference is that the structure~\eqref{eq:G} of the cascade is not captured by this parametrization, and we now have a standard MIMO OE problem.
Although this problem also requires minimizing a non-convex cost function, standard methods are available to initialize it.
Second, using $G(q,\theta)=G(q,\eta)$, let an estimate of $G_2(q,\hat{\theta}_2)$ be given by $G_{12}(q,\hat{\eta}_{12})$, and consider
\begin{equation}
\hat{G}_1(q) = \frac{G_{11}(q,\hat{\eta}_{11})}{G_{12}(q,\hat{\eta}_{12})} , \quad \hat{G}_3(q) = \frac{G_{11}(q,\hat{\eta}_{11})}{G_{13}(q,\hat{\eta}_{13})}
\label{eq:red}
\end{equation}
as estimates of $G_1(q)$ and $G_3(q)$.
Except for $G_2(q)$, these estimates do not have the desired structure (i.e., they are overparametrized, due to the noise in the estimates).
Third, apply a model order reduction technique to $\hat{G}_1(q)$ and $\hat{G}_3(q)$ in order to obtain estimates with the desired structures $G_1(q,\hat{\theta}_1)$ and $G_3(q,\hat{\theta}_3)$.

Although this procedure provides initial estimates for~\eqref{eq:Vpem} by solving standard PEM problems, the risk of converging to non-global minima in some cost function is still high, as the problems are MIMO.
As basis to propose a method for the cascade problem that does not have this limitation, we now consider WNSF for SISO OE models.

%% file: wnsf_siso.tex
The weighted null-space fitting (WNSF) method was introduced by Galrinho et al. (2014) and is consistent and asymptotically efficient under similar assumptions as PEM (Galrinho et al., 2017).
We now review the method, which we will extend for cascade networks.

Consider an OE model as~\eqref{eq:mimocascade} with a SISO transfer function $G(q,\theta)$.
WNSF is a three-step weighted least-squares method to estimate $\theta$.
First, a non-parametric model is estimated by least squares.
Second, the estimated non-parametric model is reduced to a parametric model by least squares.
Third, the parametric model is re-estimated by weighted least squares, where the weighting is constructed using the parametric estimate obtained in Step 2.

\subsubsection{Step 1: Non-parametric model}
In the first step, we approximate the parametric OE model with the non-parametric FIR model $y(t) = \sum_{k=0}^{n-1} g_k q^{-k} u(t) + e(t)$, where $n$ is sufficiently large so that the bias error by truncation is negligible.
The PEM estimate of $g^n = 
\begin{bmatrix}
g_0 & g_1 & \cdots & g_{n-1}
\end{bmatrix}^\top$
is then obtained by least squares with
\begin{equation}
\hat{g}^n = \left[\frac{1}{N}\sum_{t=1}^N \varphi(t)\varphi^\top(t)\right]^{-1}
\left[ \frac{1}{N}\sum_{t=1}^N \varphi(t) y(t) \right],
\label{eq:gls}
\end{equation}
where
$\varphi(t) = 
\begin{bmatrix}
u(t) & u(t-1) & \cdots & u(t-n-1)
\end{bmatrix}^\top .$
If we assume the error made by truncation is negligible, the noise in the non-parametric estimate $\Delta_g^n = \hat{g}^n-g_\nul^n$ ($g_\nul$ are the first $n$ true impulse response coefficients) is distributed as
$\Delta_g^n \sim \mathcal{N} (0, P)$, where $\mathcal{N}$ is the normal distribution and
\begin{equation}
P = \left[\sum_{t=1}^N \varphi(t) \frac{1}{\lambda} \varphi^\top(t)\right]^{-1}.
\label{eq:P}
\end{equation}

\subsubsection{Step 2: Estimation of Parametric Model}
In the second step, we use the relation
\begin{equation}
\frac{B(q,\theta)}{F(q,\theta)} = \sum_{k=0}^{n-1} g_k q^{-k}
\label{eq:B/F=G}
\end{equation}
to obtain an estimate of $\theta$ from the estimate of $g^n$ we have obtained in Step 1.
This is done by re-writing~\eqref{eq:B/F=G} as
\begin{equation}
F(q,\theta) \sum_{k=0}^n g^n_k q^{-k} - B(q,\theta) = 0.
\end{equation}
This can be re-written in matrix form as 
\begin{equation}
g^n - Q(g^n)\theta = 0,
\label{eq:g=Qtheta}
\end{equation}
where
\begin{equation}
Q(g^n) =
\begin{bmatrix}
-\mathcal{T}_{n\times m}\{\Gamma_n g_{12}^n\} & \begin{matrix}
I_{m\times m} \\ 0_{n-m\times m}
\end{matrix}
\end{bmatrix},
\end{equation}
with
\begin{equation}
\Gamma_n =
\begin{bmatrix}
0_{1\times n-1} & 0 \\
I_n & 0_{n-1\times 1}
\end{bmatrix}.
\end{equation}
Because~\eqref{eq:g=Qtheta} is linear in $\theta$, we may replace $g^n$ by the estimate $\hat{g}^n$ from Step 1, and solve for $\theta$ with least squares:
\begin{equation}
\hat{\theta}_\text{LS} = \left[ Q^\top(\hat{g}^n) Q(\hat{g}^n) \right]^{-1} Q^\top(\hat{g}^n) \hat{g}^n .
\label{eq:thetals}
\end{equation}

\subsubsection{Step 3: Re-estimation of Parametric Model}
The idea of the third step is to re-estimate $\theta$ with a statistically sound approach.
We do this by replacing the noisy estimate $\hat{g}^n$ in~\eqref{eq:g=Qtheta} and re-writing the residuals as
\begin{equation}
\begin{aligned}
\hat{g}^n - Q(\hat{g}^n) \theta &= g^n+\Delta^n_g - Q(g^n+\Delta^n_g) \theta  \\
&= [g^n-Q(g^n)\theta] + [\Delta^n_g - \tilde{Q}(\Delta^n_g) \theta] ,
\end{aligned}
\label{eq:res}
\end{equation}
where $\tilde{Q}$ is defined similarly to $Q$ but with the identity matrix in the top-right block replaced by the zero matrix.
Using~\eqref{eq:g=Qtheta}, \eqref{eq:res} can be re-written as
\begin{equation}
\hat{g}^n - Q(\hat{g}^n) \theta = T(\theta) \Delta^n_g,
\label{eq:TDelta}
\end{equation}
where $T(\theta)=\mathcal{T}_{n\times n}\{ \begin{bmatrix}
1 & f_1 & \dots & f_m
\end{bmatrix}^\top \}$.
If the residuals we try to minimize when we replace $g^n$ by $\hat{g}^n$ in~\eqref{eq:g=Qtheta} are given by~\eqref{eq:TDelta}, the estimate of $\theta$ with minimum variance is given by solving a weighted least-squares problem, where the weighting is the inverse of the covariance of the residuals~\eqref{eq:TDelta}.
This covariance is given by
\begin{equation}
W^{-1}(\theta) = T(\theta) P T^{-1}(\theta) .
\label{eq:Winv}
\end{equation}
Because the true value of $\theta$ is not available to compute~\eqref{eq:Winv}, we replace it by the estimate $\hat{\theta}_\text{LS}$ obtained in Step 2. 
The noise variance $\lambda$ in $P$~\eqref{eq:P} is also typically unknown, but because it is a scalar, it can be disregarded in the weighting.
Then, we re-estimate $\theta$ using
\begin{equation}
\hat{\theta}_\text{WLS} = \left[ Q^\top(\hat{g}^n) W(\hat{\theta}_\text{LS}) Q(\hat{g}^n) \right]^{-1} Q^\top(\hat{g}^n) W(\hat{\theta}_\text{LS}) \hat{g}^n .
\label{eq:thetawls}
\end{equation}

The estimate $\hat{\theta}_\text{WLS}$ is an asymptotically efficient estimate of $\theta$.
However, for finite sample size, continuing to iterate, constructing the weighting with the estimate obtained at the previous iteration, may provide an improvement.

\textbf{Remark 2.}
Three properties are required to apply WNSF:
\begin{enumerate}
\item the parametric model of interest can be approximated by a non-parametric model estimated by least squares;
\item the non-parametric and parametric models can be related using the form~\eqref{eq:g=Qtheta} (i.e., linear in the parametric model parameters);
\item the residuals~\eqref{eq:TDelta} are linear in the error of the non-parametric estimate, so that a closed-form expression for the covariance can be obtained. 
\end{enumerate}

%% file: wnsf_cascade.tex
In this section, we extend WNSF for cascade networks, using the network in Fig.~\ref{fig:network1} as example.
We follow the WNSF steps from Section~\ref{sec:wnsf}.

\subsubsection{Step 1: Non-parametric model}
The key aspect of Step 1, as stated in Remark 2, is to estimate with least squares a non-parametric model that can approximate the parametric model of interest.
For the cascade network in Fig.~\ref{fig:network1}, we use the non-parametric MIMO FIR model
\begin{equation}
\begin{bmatrix}
y_1(t) \\ y_2(t)
\end{bmatrix} =
\begin{bmatrix}
\tilde{G}_{11}(q,g^n_{11}) & \tilde{G}_{12}(q,g^n_{12}) \\
\tilde{G}_{21}(q,g^n_{21}) & \tilde{G}_{22}(q,g^n_{22}) 
\end{bmatrix}
u(t) +
\begin{bmatrix}
e_1(t) \\ e_2(t)
\end{bmatrix},
\label{eq:mimofir}
\end{equation}
where, for $i,j=\{1,2\}$,
$\tilde{G}_{ij}(q,g^n_{ij}) = \sum_{k=0}^{n-1} g^{(ij)}_k q^{-k}$, and
$
g^n_{ij} =
[
g^{(ij)}_0 \; g^{(ij)}_1 \; \cdots \; g^{(ij)}_{n-1}
]^\top.
$
Then, the PEM estimate of 
$
g^n =
\begin{bmatrix}
(g^n_{11})^\top & (g^n_{12})^\top & (g^n_{21})^\top (g^n_{22})^\top
\end{bmatrix}^\top
$
is obtained by least squares with~\eqref{eq:gls}, but with $y(t) = 
\begin{bmatrix}
y_1(t) & y_2(t) 
\end{bmatrix}^\top$ and
\begin{equation}
\varphi(t) =
\begin{bmatrix}
\varphi_{11}(t) & 0 \\
\varphi_{12}(t) & 0 \\
0 & \varphi_{21}(t) \\
0 & \varphi_{22}(t)
\end{bmatrix},
\end{equation}
where
\begin{equation}
\begin{aligned}
\varphi_{11}(t) =
&\begin{bmatrix}
u_1(t) & u_1(t-1) & \cdots & u_1(t-n-1) \\
\end{bmatrix}^\top = \varphi_{21}(t),\\
\varphi_{12}(t) =
&\begin{bmatrix}
u_2(t) & u_2(t-1) & \cdots & u_2(t-n-1) \\
\end{bmatrix}^\top = \varphi_{22}(t) .
\end{aligned}
\label{eq:phis}
\end{equation}
Now, the covariance of $\hat{g}^n$ is given by, similarly to~\eqref{eq:P} but for the MIMO case,
\begin{equation}
P = \left[\sum_{t=1}^N \varphi(t) \Lambda^{-1} \varphi^\top(t)\right]^{-1},
\label{eq:Pmimo}
\end{equation}
where $\Lambda$ is a diagonal matrix with diagonal $[\lambda_1 \; \lambda_2]$.

\subsubsection{Step 2: Estimation of Parametric Model}
The second step consists of estimating $\theta$ from the non-parametric estimate.
The key aspect, as mentioned in Remark 2, is that the relation between the non-parametric and the parametric model can be written linearly in the parameters $\theta$, as in~\eqref{eq:g=Qtheta}.
In this case, we have
\begin{equation}
\begin{bmatrix}
G_2^\theta G_1^\theta & G_2^\theta\\
G_3^\theta G_2^\theta G_1^\theta & G_3^\theta G_2^\theta
\end{bmatrix}
=
\begin{bmatrix}
\tilde{G}_{11}^g & \tilde{G}_{12}^g \\
\tilde{G}_{21}^g & \tilde{G}_{22}^g
\end{bmatrix} ,
\label{eq:G=G}
\end{equation}
where we used the simplified notation $G_j^\theta=G_j(q,\theta_j)$ and $\tilde{G}^g_{ij}=\tilde{G}(q,g^n_{ij})$. 
Because of the products between transfer functions parametrized by $\theta$, this cannot be written linearly in $\theta$ as in~\eqref{eq:g=Qtheta}.
Thus, the second point in Remark 2 is not satisfied, and WNSF cannot be applied directly.

To solve this problem, we re-write~\eqref{eq:G=G} as
\begin{equation}
\begin{bmatrix}
G_1^\theta \tilde{G}_{12}^g & G_2^\theta\\
G_1^\theta \tilde{G}_{22}^g & G_3^\theta \tilde{G}_{12}^g
\end{bmatrix}
=
\begin{bmatrix}
\tilde{G}_{11}^g & \tilde{G}_{12}^g \\
\tilde{G}_{21}^g & \tilde{G}_{22}^g
\end{bmatrix} ,
\label{eq:G=Grep}
\end{equation}
where we replaced some of the products of $\theta$ by non-parametric models.
Replacing $G_j^\theta$ with the respective numerator and denominator, we can re-write~\eqref{eq:G=Grep} as
\begin{equation}
\begin{bmatrix}
F_1^\theta \tilde{G}_{11}^g-L_1^\theta\tilde{G}_{12}^g & F_2^\theta \tilde{G}_{12}^g-L_2^\theta \\
F_1^\theta \tilde{G}_{21}^g-L_1^\theta \tilde{G}_{22}^g & F_3^\theta \tilde{G}_{22}^g - L_3^\theta\tilde{G}_{12}^g
\end{bmatrix} = 0,
\label{eq:FG-LG=0}
\end{equation}
which is linear in $\theta$, and can be written as~\eqref{eq:g=Qtheta} with 
\begin{equation}
Q(g^n) =
\begin{bmatrix}
Q_1(g^n) & 0 & 0 \\
0 & Q_2(g^n) & 0 \\
Q_3(g^n) & 0 & 0 \\
0 & 0 & Q_4(g^n)
\end{bmatrix},
\end{equation}
where
\begin{equation}
\begin{aligned}
Q_1(g^n) &= 
\begin{bmatrix}
-\mathcal{T}_{n\times m}\{\Gamma_n g_{11}^n\} & \mathcal{T}_{n\times m}\{g_{12}^n\}
\end{bmatrix}, \\
Q_2(g^n) &= 
\begin{bmatrix}
-\mathcal{T}_{n\times m}\{\Gamma_n g_{12}^n\} & \begin{matrix}
I_{m\times m} \\ 0_{n-m\times m}
\end{matrix}
\end{bmatrix}, \\
Q_3(g^n) &= 
\begin{bmatrix}
-\mathcal{T}_{n\times m}\{\Gamma_n g_{21}^n\} & \mathcal{T}_{n\times m}\{g_{22}^n\}
\end{bmatrix}, \\
Q_4(g^n) &= 
\begin{bmatrix}
-\mathcal{T}_{n\times m}\{\Gamma_n g_{22}^n\} & \mathcal{T}_{n\times m}\{g_{12}^n\}
\end{bmatrix}.
\end{aligned}
\end{equation}
Then, the estimate $\hat{g}^n$ from Step 1 can be used to obtain an estimate of $\hat\theta_\text{LS}$ with least squares, using~\eqref{eq:thetals}.

\textbf{Remark 3.}
The replacement from~\eqref{eq:G=G} to~\eqref{eq:G=Grep} to achieve linearity in $\theta$ is not unique.
For example, 
\begin{equation}
\begin{bmatrix}
G_1^\theta \tilde{G}_{12}^g & G_2^\theta\\
G_3^\theta \tilde{G}_{11}^g & G_3^\theta \tilde{G}_{12}^g
\end{bmatrix}
=
\begin{bmatrix}
\tilde{G}_{11}^g & \tilde{G}_{12}^g \\
\tilde{G}_{21}^g & \tilde{G}_{22}^g
\end{bmatrix}
\label{eq:G=Grep2}
\end{equation}
is also possible.
As we will see in Section~\ref{sec:sim}, simulations support that the replacement chosen does not change the asymptotic properties of the estimate.

\textbf{Remark 4.}
Although not all equations defined in~\eqref{eq:FG-LG=0} need to be used to obtain consistent estimates of all transfer functions (in this case, there are two equations that determine $G_1^\theta$), discarding ``redundant'' equations will have a cost in terms of efficiency.

\subsubsection{Step 3: Re-estimation of Parametric Model}
In the third step, we re-estimate $\theta$ by weighted least squares.
The key aspect to derive the weighting matrix, as mentioned in Remark 2, is that the residuals of~\eqref{eq:FG-LG=0} when $\tilde{G}_{ij}^g$ are replaced by estimates can be written linearly in the non-parametric estimate noise, as in~\eqref{eq:TDelta}.
This will be possible because~\eqref{eq:FG-LG=0} can be written linearly in $g^n$.
In this case, $T(\theta)$ is given by
\begin{equation}
T(\theta) =
\begin{bmatrix}
T_{11}(\theta) & 0_{2n\times 2n} \\
T_{21}(\theta) & T_{22}(\theta)
\end{bmatrix},
\end{equation}
where
\begin{equation}
\begin{aligned}
T_{11}(\theta) &= 
\begin{bmatrix}
\mathcal{T}_{n\times n} \{ f^{(1)} \} &
-\mathcal{T}_{n\times n} \{ b^{(1)} \} \\
0_{n\times n} & \mathcal{T}_{n\times n} \{ f^{(2)} \}
\end{bmatrix}, \\
T_{21}(\theta) &= 
\begin{bmatrix}
0_{n\times n} & 0_{n\times n} \\
0_{n\times n} & -\mathcal{T}_{n\times n}\{ b^{(3)} \} \}
\end{bmatrix}, \\
T_{22}(\theta) &= 
\begin{bmatrix}
\mathcal{T}_{n\times n} \{ f^{(1)} \} &
-\mathcal{T}_{n\times n} \{ b^{(1)} \} \\
0_{n\times n} & \mathcal{T}_{n\times n} \{ f^{(3)} \}
\end{bmatrix},
\end{aligned}
\end{equation}
with $f^{(j)} = [1 \; f_1^{(j)} \; \dots \; f_m^{(j)}]^\top$ and $b^{(j)} = [b_0^{(j)} \; \dots \; b_{m-1}^{(j)}]^\top$, while the covariance of $\Delta_g^n$ is given by~\eqref{eq:Pmimo}.
In the SISO case, it was given by~\eqref{eq:P}, where $\lambda$ was a scalar and could be neglected in the weighting.
Now, this is replaced by $\Lambda$, which is a matrix.
Thus, it cannot be ignored in the weighting and must be estimated from data using
$
\hat{\Lambda} =
\frac{1}{N} \sum_{t=1}^N \varepsilon(t,\hat{\theta}_\text{LS}) \varepsilon^\top(t,\hat{\theta}_\text{LS}),
$
where the off-diagonal elements can then be replaced with zeros under the assumption that the noise sources are mutually independent. Then, if we denote by $\hat{P}$ the matrix~\eqref{eq:Pmimo} with $\Lambda$ replaced by $\hat{\Lambda}$, we can re-estimate $\theta$ with weighted least squares using~\eqref{eq:thetawls}, where
\begin{equation}
W^{-1}(\theta) = T(\theta) \hat P T^\top(\theta) .
\label{eq:WinvPhat}
\end{equation}

Because of the similarities with the SISO case and the theoretical analysis by~\citet{wnsf_journal}, there are reasons to believe that $\hat{\theta}_\text{WLS}$ is an asymptotically efficient estimate of $\theta$.
Also as in the SISO case, for finite sample size, it may be possible to improve the accuracy by iterating, where the weighting at some iteration is constructed using the estimate from the previous iteration.

\textbf{Algorithm 1}
The following WNSF algorithm can be applied to identify the cascade network in Fig.~\ref{fig:network1}:
\begin{enumerate}
\item estimate the non-parametric FIR model~\eqref{eq:mimofir} with least squares~\eqref{eq:gls};
\item estimate the parametric model with least squares~\eqref{eq:thetals}, using~\eqref{eq:FG-LG=0} or~\eqref{eq:G=Grep2} with the non-parametric estimate;
\item re-estimate the parametric model with weighted least squares~\eqref{eq:thetawls}, with~\eqref{eq:WinvPhat} the weighting inverse.
\end{enumerate}

%% file: simulations.tex
In this section, we perform a simulation study to illustrate the performance of the method.
Mainly, we observe how the simulation supports that the method is asymptotically efficient, and how it is robust against convergence to non-global minima of the PEM cost function.

In the simulation, we use the network in Fig.~\ref{fig:network1} with:
\begin{equation}
\begin{aligned}
G_1(q)\! &= \! \frac{0.7 q^{-1}+0.5q^{-2}}{1\!-\!1.2q^{-1}\!+\!0.5q^{-2}}, \;
G_2(q) \! = \! \frac{0.6-0.2q^{-1}}{1\!-\!1.3q^{-1}\!+\!0.6q^{-2}}, \\
G_3(q) \! &= \! \frac{0.6+0.8q^{-1}-1.2q^{-2}}{1-0.75q^{-1}+0.56q^{-2}}.
\end{aligned}
\end{equation}
The noises $\{e_1(t),e_2(t)\}$ have variances 2 and 3, respectively, and the inputs $\{u_1(t),u_2(t)\}$ are given by $u_j(t)=(1-0.9q^{-1})^{-1} e_j^u(t)$, where $\{e_1^u(t),e_2^u(t)\}$ are unit-variance mutually-uncorrelated Gaussian white-noise sequences.

We compare the following methods:
\begin{itemize}
\item minimization of the cost function~\eqref{eq:Vpem} initialized at the true parameter values (PEM-true);
\item minimization of the cost function~\eqref{eq:Vpem} initialized according to the procedure described in Section 3 (PEM-oe);
\item weighted null-space fitting according to the relations~\eqref{eq:FG-LG=0} (denoted WNSF-1);
\item weighted null-space fitting according to the relations~\eqref{eq:G=Grep2} (denoted WNSF-3).
\end{itemize}
For the PEM methods, the network structure is implemented with the MATLAB function \texttt{greyest}, and then minimized using the \texttt{pem} function, with a maximum of 1000 iterations.
For the initialization of PEM-oe, we first estimate the over-parametrized MIMO OE model~\eqref{eq:mimo_oe_overp} using the MATLAB function \texttt{oe} with default initialization.
Then, an estimate of $G_2(q,\theta)$ is readily available through $G_{12}(q,\hat\eta_{12})$, while estimates of $G_1(q,\theta)$ and $G_3(q,\theta)$ are obtained by performing model order reduction on the over-parametrized estimates~\eqref{eq:red}; for this, we use the MATLAB function \texttt{reduce}.
For WNSF, we apply the algorithm for a grid of non-parametric model orders $n=\{20,30,40\}$ (all the polynomials in the MIMO FIR were chosen to have the same order) and a maximum of 1000 iterations, and the parametric estimate that minimizes the PEM criterion~\eqref{eq:Vpem} is chosen.
This is the same approach that has been used for the SISO case~\citep{wnsf_journal}.

\begin{figure}
\centering
\input{figMSE.tex}
\vspace{-1em}
\caption{Average MSE as function of sample size.}
\label{fig:MSE}
\end{figure}
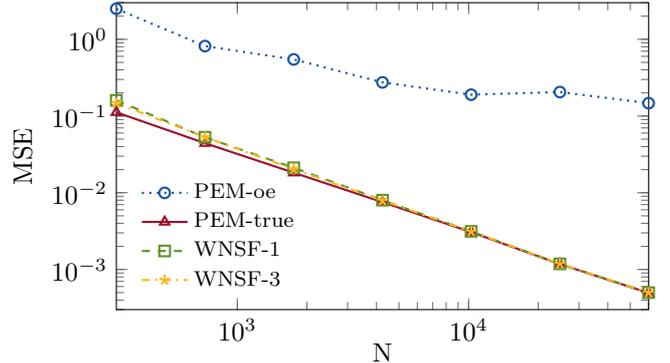

We perform 1000 Monte Carlo runs for seven sample sizes $N$ logarithmically spaced between 300 and 60000 (rounded to the nearest integer).
The results are presented in Fig.~\ref{fig:MSE}, where we plot the average mean-square error (MSE$=||\hat{\theta}-\theta_\nul||^2$) as function of the sample size $N$, where $\hat\theta$ is the estimate obtained for a particular method and sample size.
In Table 1, we present the average computational times for each method and sample size $N$.

\begin{table}
  \caption{Average computational times (in seconds) for several sample sizes.}
  \label{tbl:computationaltimes}
  \vspace{-2mm}
  \begin{center}
    \begin{tabular}{ l | *{7}{c} }
      \multicolumn{1}{r|}{$N$} & 300 & 725 & 1754 & 4243 & 10260 & 24811 & 60000 \\
      \hline
      PEM-oe  & 40 & 24 & 19   & 18  & 18 & 26  & 41 \\
      PEM-true  & 11 & 10  & 9.1   & 10  & 12 & 18 & 31 \\
      WNSF-1   & 5.0 & 2.4 & 1.7   & 1.6  & 1.8 & 3.0 & 6.1 \\
      WNSF-3  & 4.7 & 2.2 &  1.6    & 1.5  & 1.8 & 2.9 & 6.1 \\
      \hline
    \end{tabular}
  \end{center}
\end{table}

We draw the following conclusions from these results.
First, the two alternative procedures~\eqref{eq:G=Grep} and~\eqref{eq:G=Grep2} to make the equations for WNSF linear perform similarly.
Second, initializing PEM by estimating an over-parametrized MIMO OE model followed by model reduction often did not provide accurate enough estimates to attain the global minimum of the cost function, as the MSE is considerably different than when PEM is initialized at the true parameters.
Third, WNSF is an appropriate method to avoid the non-convexity of PEM, as it did not converge to low-performance non-global minima.
Fourth, the computational time was considerably lower for WNSF than for PEM, even when initialized at the true parameters (for PEM-oe the times in Table~\ref{tbl:computationaltimes} do not take into account the MIMO OE estimate for initialization).
Fifth, as supported by the SISO analysis by~\citet{wnsf_journal}, the delineated network WNSF procedure seems to be asymptotically efficient.

%% file: figMSE.tex
\begin{tikzpicture}
  \begin{axis}[%
    width=\figurewidth,
    height=\figureheight,
    at={(0.758in,0.481in)},
    scale only axis,
    xmode=log,
    xmin=300,
    xmax=60000,
    xtick = {1000,10000},
    xlabel={N},
    xlabel style={yshift=2mm},
    ymode=log,
    ymin=.0003,
    ymax=3,
    yminorticks=true,
    ylabel={MSE},
    axis background/.style={fill=white},
    every axis plot/.append style={semithick},
    legend style={at={(0.03,0.03)},anchor=south west,legend cell align=left,align=left,draw=none,fill=none,font=\small},
    cycle list name=kth-cycle,
  ]
    \addplot+[mark=o,dotted,every mark/.append style={solid}] table[col sep=comma]
    {pem_oe.csv};
    \addlegendentry{PEM-oe};

    \addplot+[mark=triangle,solid,every mark/.append style={solid}] table[col sep=comma]{pem_true.csv};
    \addlegendentry{PEM-true};
    
    \addplot+[mark=square,dashed,every mark/.append style={solid}] table[col sep=comma]{wnsf1.csv};
    \addlegendentry{WNSF-1};

    \addplot+[mark=star,dashdotted,every mark/.append style={solid}] table[col sep=comma]{wnsf3.csv};
    \addlegendentry{WNSF-3};

  \end{axis}
\end{tikzpicture}%

%% file: extensions.tex
In this paper, we proposed a method for identification of networks with sensor noise.
The method is based on WNSF, which has been shown to be asymptotically efficient for SISO models~\citep{wnsf_journal}.
A simulation supports the idea that this extension of WNSF is still asymptotically efficient, and the aforementioned work gives theoretical support to this idea.
More thorough simulations to test robustness will be performed in the future.

For simplicity of presentation, we introduced the method for the network in Fig.~\ref{fig:network1}.
We now address how the method generalizes to other networks.
In particular, serial cascade networks for which all nodes with an input appear prior to all nodes with outputs are straightforwardly covered by the proposed algorithm.
For space concerns, this case is not detailed here in general, but it will be considered in a future contribution.
Moreover, if additional inputs or outputs exist, the algorithm is also applicable straightforwardly, but there will be additional equations to consider.


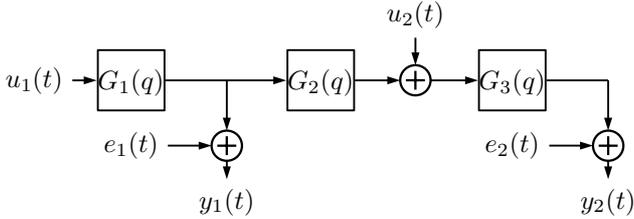
\begin{figure}
\input{network2.tex}
\vspace{-2.5em}
\caption{Another serial cascade network example.}
\label{fig:network2}
\end{figure}

Serial cascade networks that are not covered straightforwardly occur when not all nodes with inputs appear prior to all nodes with outputs.
Consider the cascade network in Fig.~\ref{fig:network2}. 
In this case, the relation between the parametric and non-parametric models would be given by
\begin{equation}
\begin{bmatrix}
G_1^\theta & 0 \\
G_3^\theta G_2^\theta G_1^\theta & G_3^\theta
\end{bmatrix}
-
\begin{bmatrix}
\tilde{G}_{11}^g & 0 \\
\tilde{G}_{21}^g & \tilde{G}_{22}^g
\end{bmatrix}
= 0.
\label{eq:impossibleG=G}
\end{equation}
Although~\eqref{eq:impossibleG=G} can be made linear in $\theta$ by replacing $G^\theta_3 G^\theta_2 G^\theta_1 = \tilde{G}_{22}^g G^\theta_2 \tilde{G}_{11}^g$, a closed-form expression for the weighting is not available, because it will not be possible to write the residuals of~\eqref{eq:impossibleG=G} linearly in the non-parametric estimate noise (the third criterion in Remark 2 is not satisfied).
This is a consequence of the product $\tilde{G}_{11}^g\tilde{G}_{22}^g$.
Solving this requires an intermediate step, which will be detailed in a future contribution.

If the measurement noise is colored (and possibly correlated between the different outputs), a non-parametric ARX model should be estimated instead of an FIR, as in the SISO case~\citep{Galrinho:14}.
However, differently than the SISO case, products of non-parametric ARX polynomials will now always appear in Step 2, similarly to~\eqref{eq:impossibleG=G}.
This hinders successful application of Step 3, and solving it requires a similar approach as to solve the aforementioned case, which will be detailed in the same contribution.

These contributions will complete the proposed algorithm for serial cascade networks.
However, the same principles can be used to apply WNSF to other network structures affected by sensor noise.
These contributions will then serve as basis to propose a more general network WNSF method.


%% file: network2.tex
\begin{tikzpicture}

	\matrix (m1) [row sep=0.8mm, column sep=3mm]
	{
		\node[coordinate]  (m00) {};     &
		\node[coordinate]          (m01) {};          &
		\node[coordinate]         (m02) {};  &
		\node[coordinate]                 (m03) {};          &
        \node[dsp/label=above]                  (m04) {$u_2(t)$};    &
		\node[coordinate]     (m05) {};  &
		\node[coordinate]                 (m06) {};\\
        \\
		\node[dsp/label=left]  (m10) {$u_1(t)$};     &
		\node[dspsquare]                  (m11) {$G_1(q)$};          &
		\node[coordinate]      (m12) {};  &
		\node[dspsquare]   (m13) {$G_2(q)$};          &
        \node[dspadder]                  (m14) {};          &
		\node[dspsquare]     (m15) {$G_3(q)$};  &
		\node[coordinate]  (m16) {};    \\
        \\
		\node[coordinate]  (m20) {};     &
		\node[dsp/label=left]                  (m21) {$e_1(t)$};          &
		\node[dspadder]      (m22) {};  &
		\node[coordinate]   (m23) {};          &
        \node[coordinate]   (m24) {};          &
		\node[dsp/label=left]     (m25) {$e_2(t)$};  &
		\node[dspadder]  (m26) {};    \\
        \\
		\node[coordinate]  (m30) {};     &
		\node[coordinate]                  (m31) {};          &
		\node[dsp/label=below]      (m32) {$y_1(t)$};  &
		\node[coordinate]   (m33) {};          &
        \node[coordinate]                  (m34) {};          &
		\node[coordinate]     (m35) {};  &
		\node[dsp/label=below]  (m36) {$y_2(t)$};     \\
        \\
	};

    
    \begin{scope}[start chain]
		\chainin (m10);
		\chainin (m11) [join=by dspconn];
	\end{scope}
    \begin{scope}[start chain]
		\chainin (m11);
		\chainin (m13) [join=by dspconn];
	\end{scope}
    \begin{scope}[start chain]
		\chainin (m04);
		\chainin (m14) [join=by dspconn];
	\end{scope}
    \begin{scope}[start chain]
		\chainin (m13);
        \chainin (m14) [join=by dspconn];
		\chainin (m15) [join=by dspconn];
	\end{scope}
    \begin{scope}[start chain]
		\chainin (m12);
		\chainin (m22) [join=by dspconn];
	\end{scope}
    \begin{scope}[start chain]
    	\chainin (m15);
		\chainin (m16) [join=by dspline];
        \chainin (m26) [join=by dspconn];
	\end{scope}
    \begin{scope}[start chain]
		\chainin (m21);
		\chainin (m22) [join=by dspconn];
	\end{scope}
        \begin{scope}[start chain]
		\chainin (m25);
		\chainin (m26) [join=by dspconn];
	\end{scope}
    \begin{scope}[start chain]
		\chainin (m22);
		\chainin (m32) [join=by dspconn];
	\end{scope}
    \begin{scope}[start chain]
		\chainin (m26);
		\chainin (m36) [join=by dspconn];
	\end{scope}
    

\end{tikzpicture}

%% file: ifacconf.bbl
\begin{thebibliography}{23}
\providecommand{\natexlab}[1]{#1}
\providecommand{\url}[1]{\texttt{#1}}
\providecommand{\urlprefix}{URL }
\expandafter\ifx\csname urlstyle\endcsname\relax
  \providecommand{\doi}[1]{doi:\discretionary{}{}{}#1}\else
  \providecommand{\doi}{doi:\discretionary{}{}{}\begingroup
  \urlstyle{rm}\Url}\fi

\bibitem[{Dankers et~al.(2015)Dankers, Van~den Hof, Bombois, and
  Heuberger}]{Dankers:15EIV}
Dankers, A., Van~den Hof, P., Bombois, X., and Heuberger, P. (2015).
\newblock Errors-in-variables identification in dynamic networks---consistency
  results for an instrumental variable approach.
\newblock \emph{Automatica}, 62, 39--50.

\bibitem[{Dankers et~al.(2016)Dankers, Van~den Hof, Bombois, and
  Heuberger}]{dankers2016identification}
Dankers, A., Van~den Hof, P.M., Bombois, X., and Heuberger, P.S. (2016).
\newblock Identification of dynamic models in complex networks with prediction
  error methods: Predictor input selection.
\newblock \emph{IEEE Transactions on Automatic Control}, 61(4), 937--952.

\bibitem[{Everitt et~al.(2016)Everitt, Bottegal, Rojas, and
  Hjalmarsson}]{everitt2016identification}
Everitt, N., Bottegal, G., Rojas, C.R., and Hjalmarsson, H. (2016).
\newblock Identification of modules in dynamic networks: An empirical bayes
  approach.
\newblock In \emph{55th IEEE Conference on Decision and Control}, 4612--4617.

\bibitem[{Everitt et~al.(2013)Everitt, Rojas, and Hjalmarsson}]{Niklas:13}
Everitt, N., Rojas, C.R., and Hjalmarsson, H. (2013).
\newblock A geometric approach to variance analysis of cascaded systems.
\newblock In \emph{52nd IEEE Conference on Decision and Control}, 6496--6501.

\bibitem[{Everitt et~al.(2014)Everitt, Rojas, and Hjalmarsson}]{Niklas:14}
Everitt, N., Rojas, C.R., and Hjalmarsson, H. (2014).
\newblock Variance results for parallel cascade serial systems.
\newblock In \emph{19th IFAC World Congress}, 2317--2322.

\bibitem[{Forssell and Ljung(1999)}]{Forssell:99}
Forssell, U. and Ljung, L. (1999).
\newblock Closed-loop identification revisited.
\newblock \emph{Automatica}, 35(7), 1215--1241.

\bibitem[{Galrinho et~al.(2017{\natexlab{a}})Galrinho, Rojas, and
  Hjalmarsson}]{wnsf_journal}
Galrinho, M., Rojas, C.R., and Hjalmarsson, H. (2017{\natexlab{a}}).
\newblock Parametric identification with weighted null-space fitting.
\newblock \emph{submitted to Transactions on Automatic Control
  (arXiv:1708.03946)}.

\bibitem[{Galrinho et~al.(2017{\natexlab{b}})Galrinho, Everitt, and
  Hjalmarsson}]{network_morsm}
Galrinho, M., Everitt, N., and Hjalmarsson, H. (2017{\natexlab{b}}).
\newblock Incorporating noise modeling in dynamic networks using non-parametric
  models.
\newblock In \emph{20th IFAC World Congress}, 4612--4617.

\bibitem[{Galrinho et~al.(2014)Galrinho, Rojas, and Hjalmarsson}]{Galrinho:14}
Galrinho, M., Rojas, C.R., and Hjalmarsson, H. (2014).
\newblock A weighted least-squares method for parameter estimation in
  structured models.
\newblock In \emph{53rd IEEE Conference on Decision and Control}, 3322--3327.

\bibitem[{Gevers et~al.(2017)Gevers, Bazanella, and
  Parraga}]{gevers2017identifiability}
Gevers, M., Bazanella, A.S., and Parraga, A. (2017).
\newblock On the identifiability of dynamical networks.
\newblock In \emph{20th IFAC World Congress}, 10580--10585.

\bibitem[{H{\"a}gg et~al.(2010)H{\"a}gg, Wahlberg, and
  Sandberg}]{hagg2010subspace}
H{\"a}gg, P., Wahlberg, B., and Sandberg, H. (2010).
\newblock On subspace identification of cascade structured systems.
\newblock In \emph{49th IEEE Conf. on Decision and Control}, 2843--2848.

\bibitem[{Ljung(1999)}]{ljung99}
Ljung, L. (1999).
\newblock \emph{{System Identification. Theory for the User}}.
\newblock Prentice-Hall.

\bibitem[{Materassi and Innocenti(2010)}]{materassi2010topological}
Materassi, D. and Innocenti, G. (2010).
\newblock Topological identification in networks of dynamical systems.
\newblock \emph{IEEE Transactions on Automatic Control}, 55(8), 1860--1871.

\bibitem[{Materassi and Salapaka(2012)}]{materassi2012problem}
Materassi, D. and Salapaka, M.V. (2012).
\newblock On the problem of reconstructing an unknown topology via locality
  properties of the wiener filter.
\newblock \emph{IEEE Transactions on Automatic Control}, 57(7), 1765--1777.

\bibitem[{S\"{o}derstr\"{o}m and Stoica(1983)}]{ssiv}
S\"{o}derstr\"{o}m, T. and Stoica, P. (1983).
\newblock \emph{{Instrumental Variable Methods for System Identification}}.
\newblock Springer.

\bibitem[{S{\"o}derstr{\"o}m et~al.(1991)S{\"o}derstr{\"o}m, Stoica, and
  Friedlander}]{soderstrom1991indirect}
S{\"o}derstr{\"o}m, T., Stoica, P., and Friedlander, B. (1991).
\newblock An indirect prediction error method for system identification.
\newblock \emph{Automatica}, 27(1), 183--188.

\bibitem[{Van~den Hof and Schrama(1993)}]{2stage}
Van~den Hof, P. and Schrama, R. (1993).
\newblock An indirect method for transfer function estimation from closed loop
  data.
\newblock \emph{Automatica}, 29(6), 1523--1527.

\bibitem[{Van~den Hof et~al.(2013)Van~den Hof, Dankers, Heuberger, and
  Bombois}]{van2013identification}
Van~den Hof, P.M., Dankers, A., Heuberger, P.S., and Bombois, X. (2013).
\newblock Identification of dynamic models in complex networks with prediction
  error methods---basic methods for consistent module estimates.
\newblock \emph{Automatica}, 49(10), 2994--3006.

\bibitem[{Van~Overschee and De~Moor(2012)}]{van2012subspace}
Van~Overschee, P. and De~Moor, B. (2012).
\newblock \emph{Subspace identification for linear systems: Theory,
  Implementation, Applications}.
\newblock Springer Science \& Business Media.

\bibitem[{Wahlberg et~al.(2009)Wahlberg, Hjalmarsson, and
  M{\aa}rtensson}]{wahlberg2009variance}
Wahlberg, B., Hjalmarsson, H., and M{\aa}rtensson, J. (2009).
\newblock Variance results for identification of cascade systems.
\newblock \emph{Automatica}, 45(6), 1443--1448.

\bibitem[{Wahlberg and Sandberg(2008)}]{wahlberg2008cascade}
Wahlberg, B. and Sandberg, H. (2008).
\newblock Cascade structural model approximation of identified state space
  models.
\newblock In \emph{47th IEEE Conf. on Decision and Control}, 4198--4203.

\bibitem[{Weerts et~al.(2015)Weerts, Dankers, and Van~den
  Hof}]{weerts2015identifiability}
Weerts, H.H., Dankers, A.G., and Van~den Hof, P.M. (2015).
\newblock Identifiability in dynamic network identification.
\newblock In \emph{17th IFAC Symposium on System Identification}, 1409--1414.

\bibitem[{Weerts et~al.(2016)Weerts, Van~den Hof, and
  Dankers}]{weerts2016identifiability}
Weerts, H.H., Van~den Hof, P.M., and Dankers, A.G. (2016).
\newblock Identifiability of dynamic networks with part of the nodes
  noise-free.
\newblock In \emph{12th IFAC Workshop on Adaptation and Learning in Control and
  Signal Processing}, 19--24.

\end{thebibliography}
